\documentclass[aps,preprint,epsfig,rotate]{revtex4}
\usepackage{graphicx}
\usepackage{bm}
\usepackage{epsfig}

\input epsf
\begin{document}
\title{Photodetachment cross-section of the negatively charged hydrogen ion}

 \author{Alexei M. Frolov}
 \email[E--mail address: ]{afrolov@uwo.ca}

\affiliation{Department of Applied Mathematics \\
 University of Western Ontario, London, Ontario N6H 5B7, Canada}

\date{\today}

\begin{abstract}

Photodetachment cross-section $\sigma_{ph}(p_e)$ of the negatively charged hydrogen ion H$^{-}$ is determined with the use of highly accurate variational wave functions constructed for 
this ion. Photodetachment cross-sections of the H$^{-}$ ion are also studied for very small and very large values of the photo-electron momentum $p_e$. Maximum of this cross-section and 
its location have been evaluated to high accuracy. In particular, we have found that $[\sigma_{ph}(p_e)]_{\max} \approx$ 3.8627035742 $\cdot 10^{-17}$ $cm^2$ at $p_e \approx$ 0.113206(1) 
$a.u.$ Our method is based upon the Rayleigh's formula for spherical Bessel functions.

\noindent 
10/2015, Preprint-2015-4/1 (this is the 3rd version by 25th of November 2015) [at.phys.; solar phys.], 13 pages.

\end{abstract}

\maketitle
\newpage

In this short paper we report the final results of highly accurate numerical calculations of the photodetachment cross-section of the negatively charged hydrogen ion H$^{-}$. This ion
contains two bound electrons. Absorption of the solar and/or stellar radiaton by the H$^{-}$ ion plays a central role in many astrophysical problems, including correct and accurate 
evaluations of the temperature balance at the surface of our planet \cite{Sob} - \cite{Zir}. The H$^{-}$ ion determines the absorption of infrared and visible radiation in photospheres 
of all stars, if temperatures in their photospheres are restricted between $T_{max} \approx$ 8,250 $K$ (late A-stars) and $T_{min} \approx$ 2,750 $K$ (early M-stars). In the late F, G 
and early K stars the absorption of infrared and visible radiation by the negatively charged hydrogen ions is maximal. This includes our Sun which is a star of spectral type G2. 
   
Since 1945 numerical computations of the photodetachment cross-section of the negatively charged hydrogen ion H$^{-}$ has attracted attention of a large number of scientists (see, e.g., 
\cite{Chand2} and references therein). To this moment a few hundreds of papers were published about different aspects of the photodetachment of the H$^{-}$ ion. Many of these contributions 
are related to the development of different numerical methods which allow to calculate the corresponding cross-sections (see, e.g., \cite{PR1955A} - \cite{JPB1997} and references therein). 
In this paper of a restricted volume we cannot discuss dramatic hystory of this important problem. Note only that a good introduction to the problem can be found in a number of textbooks 
(see, e.g., \cite{BS}, \cite{FFish}) and in a few recent papers (see, e.g., \cite{PRA2014} - \cite{Fro2014}) which also contain an extensive bibliography on this subject. It should be 
mentioned that overall accuracy of recent experimental works has been increased drastically \cite{PRA2014} - \cite{PRA2015}. In these papers the photodetachment cross-section of the 
negatively charged hydrogen ion H$^{-}$ has been measured to the accuracy which exceeds the accuracy of many calculations performed for this ion.   

In our earlier paper \cite{Fro2014} we tried to match the new level of experimental accuracy currently known for the photodetachment cross-section(s) of the negatively charged hydrogen 
ion H$^{-}$. However, in \cite{Fro2014} we applied the method which was accurate only for small absolute values of photo-electron momentum $p_e$, but for large values of $p_e$ our procedure 
was found to be numerically unstable. The crucial point of any accurate method developed for numerical calculations of the photodetachment cross-section of the negatively charged hydrogen ion 
H$^{-}$ is the actual formula which is applied for calculation of three-body integrals with the spherical Bessel functions. In \cite{Fro2015} we have found a successfull approach to this 
problem. However, in that paper we were restricted to the use of bound state wave functions constructed for the H$^{-}$ ion with relatively small number of basis funtions $N \le$ 350. In this 
study we solved all numerical issues completely by applying the extended arithmetical precision. Now our wave functions contain up to 4000 basis functions (exponents in the perimetric, or 
relative coordinates, see below). This allows us to determine the photodetachment cross-sections of the H$^{-}$ ion to very high accuracy. The accuracy of our cross-sections substantially 
exceeds the accuracy of the recent experimental papers performed for the H$^{-}$ ion. Formally, we can say that the photodetachment cross-section of the negatively charged hydrogen ion H$^{-}$ 
is now known to the accuracy which is sufficient for all present and future experimental needs. In reality, there are a number of corrections which must be taken into account to make accurate 
comparison with the current experimental data. Such corrections include finite mass correction, relativistic corrections to the incident and final wave functions, corrections for 
electron-electron correlations in the final state which forms after photodetachment of the H$^{-}$ ion, etc. 
          
We begin our analysis with the following formula for the photodetachment cross-section $\sigma_{ph}$ of the negatively charged hydrogen ion H$^{-}$ \cite{Fro2015}:
\begin{eqnarray}
 \sigma_{ph} = \frac{8}{3} \alpha a^2_0 \Bigl(\frac{p_e}{\frac{p^2_e}{2} + I_{n}}\Bigr) \Bigl| {\cal R}_{i \rightarrow f} \Bigr|^2 = 5.44924918
 \cdot 10^{-19} \Bigl(\frac{p_e}{\frac{p^2_e}{2} + I_{n}}\Bigr) \Bigl| {\cal R}_{i \rightarrow f} \Bigr|^2 \; \; \; cm^{2} \label{sigtt}
\end{eqnarray}
where  $\alpha = \frac{e^2}{\hbar c} = 7.2973525698 \cdot 10^{-3} \Bigl(\approx \frac{1}{137}\Bigr)$ is the dimensionless fine-structure constant, $a_0 \approx 5.2917721092 \cdot 10^{-9}$ $cm$ 
is the Bohr radius \cite{CRC}, $p_e$ is the absolute value of momentum of the photo-electron (this photo-electron is emitted during photodetachment of the H$^{-}$ ion), $I_{n} = - E({\rm H}^{-}) 
+ \frac{1}{2 n^2}$ is the ionization potential of the two-electron H$^{-}$ ion and ${\cal R}_{i \rightarrow f}$ is the photodetachment amplitude (for more detalis, see \cite{Fro2015}). The 
photodetachment amplitude is the overlap of the spatial derivatives of the incident wave function of the H$^{-}$ ion upon the two electron-nucleus coordinates and the final state wave function 
which is the product of the radial part of the total wave function $R_{n0}(r_{31})$ of the final hydrogen atom H in one of its bound $ns-$states and the wave function of the freely moving 
photo-electron. For the ground state in the final hydrogen atom we take $n = 1$ and $R_{n0}(r_{31}) = 2 \exp(- r_{31})$. If we can neglect by any interaction between this photo-electron and the
neutral hydrogen atom H, then the wave function of the free photo-electron is a plane wave, which is represented as a combination of products of the spherical Bessel functions of the first kind 
with the corresponding Legendre polynomials (see, e.g., \cite{Rose}, p.91). This formula is called the Rayleigh expansion of the plane wave \cite{Rose}. Conservation of the angular momentum 
during photodetachment leads to the selection of only one $\ell=1$-component in such an expansion. The explicit formula for the radial functions of the free particle motion is $2 p_e j_{\ell}(p_e 
r) = 2 p_e j_{\ell}(p_e r_{32})$ (see Eq.(33.10) in \cite{LLQ}). The factor 2 can be removed from this expression (it contributes to the factor $\frac83$ in the right-hand side of 
Eq.(\ref{sigtt})) and below we operate with the free-electron wave function which is of the form $p_e j_{\ell}(p_e r_{32})$ (more details can be found in \cite{Fro2014}). 

In this study the bound state wave functions of the H$^{-}$ ion are constructed in the form of the variational exponential expansion in the relative coordinates $r_{32}, r_{31}, r_{21}$
\begin{eqnarray}
  \Psi_{{\rm H}^{-}} &=& \frac12 ( 1 + \hat{P}_{12} ) \sum^{N}_{i=1} C_i \exp(-\alpha_i r_{32} - \beta_i r_{31} - \gamma_i r_{21}) \label{exp} 
\end{eqnarray}
where $r_{ij} = r_{ji}$ are the three relative coordinates which coincide with the scalar interparticle distances (see below), $C_i$ are the variational coefficients of this variational expansion 
and $\hat{P}_{12}$ is the permutation of identical particles (electrons, or particles 1 and 2). In Eq.(\ref{exp}) and everywhere below in this study the index 3 means the hydrogen nucleus which 
is assumed to be infinitely heavy. Parameters $\alpha_i, \beta_i$ and $\gamma_i$ (for $i = 1, \ldots, N$) in Eq.(\ref{exp}) are the non-linear parameters of the exponential variational expansion. 
These parameters must be varied very carefully in preliminary calculations of the total energy of the H$^{-}$ ion. For truly correlated three-particle wave function, Eq.(\ref{exp}), the explicit 
formula for the photodetachment amplitude is 
\begin{eqnarray}
 & & M_{i \rightarrow f} = \frac12 p_e \sum^{N}_{j=1} C_j \Bigl[\alpha_j \int_{0}^{+\infty} \int_{0}^{+\infty} \int^{r_{31} + r_{32}}_{\mid r_{31} - r_{32} \mid} R_{n0}(r_{31}) 
 j_{1}(p_e r_{32}) \times \nonumber \\
 & & \exp(-\alpha_j r_{32} -\beta_j r_{31} - \gamma_j r_{21}) r_{32} r_{31} r_{21} dr_{32} dr_{31} dr_{21} + \beta_j \times \label{int0} \\
 & & \int_{0}^{+\infty} \int_{0}^{+\infty} \int^{r_{31} + r_{32}}_{\mid r_{31} - r_{32} \mid} R_{n0}(r_{31}) j_{1}(p_e r_{32}) \exp(-\beta_j r_{32} -\alpha_j r_{31} - \gamma_j r_{21}) r_{32} 
 r_{31} r_{21} dr_{32} dr_{31} dr_{21} \Bigr] \nonumber
\end{eqnarray}
Direct calculations of such three-particle integrals in the relative coordinates is a difficult problem. However, such calculations can drastically be simplified with the use of three perimetric 
coordinates $u_1, u_2, u_3$ \cite{Fro2014}, \cite{Fro2015} instead of three relative coordinates. Moreover, some of the arising three-particle integrals look like singular integrals, but after a 
number of additional transformations these integrals are reduced to the regular expressions. In reality, each of such quasi-singular integrals is replaced by a sum of regular integrals. Similar 
quasi-singular integrals are called the Frullanian three-body integrals \cite{Fro2015}. Analytical and numerical calculations of the Frullanian three-body integrals are difficult, often numerically 
unstable and always slow. This is not convenient in actual compuations of a large number of matrix element and in \cite{Fro2015} we have developed another method based on the Rayleigh's formula for 
the spherical Bessel functions $j_{\ell}(p_e r)$ \cite{AS}. In the case of the $j_{1}(z) = j_{1}(p r_{32})$ Bessel function the Rayleigh's formula takes the form 
\begin{equation}
 j_{1}(z) = - \frac{d j_{0}(z)}{d z} = - \frac{d}{d z} \Bigl( \frac{\sin z}{z} \Bigr) \; \; \; . \; \; \; \label{eq1}
\end{equation}
For $z = p r_{32}$ one finds from this equation 
\begin{equation}
 j_{1}(p r_{32}) = -\frac{1}{p} \frac{\partial}{\partial r_{32}} \Bigl( j_{0}(p r_{32}) \Bigr) = -\frac{1}{p^{2}} \Bigl[\frac{\partial}{\partial r_{32}} 
 \Bigl( \frac{\sin (p r_{32})}{r_{32}} \Bigr) \Bigr] \; \; \; . \; \; \; \label{eq2}
\end{equation}
where $p = p_e$ is the absolute value of momentum of the outgoing photo-electron ${\bf p}_e$, while $r_{32}$ is the scalar distance between the hydrogen nucleus (heavy particle 3) and photo-electron 
(particle 2). The particle 1 designates another (first) electron which remains bound in the hydrogen atom H after photodetachment. The same system of notations is used everywhere below in this study.

Let us designate the wave function of the incident H$^{-}$ ion in the ground $1^1S-$state by $\Psi_{{\rm H}^{-}}(r_{32}, r_{31}, r_{21})$, while the notation $\phi_n(r_{31})$ stands for the wave 
function of the final $ns-$state of the hydrogen atom. Then we can write the following expression for the amplitude of the photodetachment $M_{i \rightarrow f}$ \cite{Fro2014}:
\begin{eqnarray}
 M_{i \rightarrow f} &=& p \int \phi_n(r_{31}) j_{1}(p r_{32}) \Phi_{{\rm H}^{-}}(r_{32}, r_{31}, r_{21}) dV \nonumber \\ 
 &=& - \int \phi_n(r_{31}) \Bigl[\frac{\partial}{\partial r_{32}} j_{0}(p r_{32}) \Bigr] \Phi_{{\rm H}^{-}}(r_{32}, r_{31}, r_{21}) dV \; \; \; . \; \; \; \label{eq3}
\end{eqnarray}  
where the notation $\Phi_{{\rm H}^{-}}(r_{32}, r_{31}, r_{21})$ is defined as follows \cite{Fro2014}
\begin{equation}
  \Phi_{{\rm H}^{-}}(r_{32}, r_{31}, r_{21}) = \Bigl( \frac{\partial}{\partial r_{32}} + \frac{\partial}{\partial r_{31}} \Bigr) \Psi_{{\rm H}^{-}}(r_{32}, r_{31}, r_{21})
\end{equation}
Also, in Eq.(\ref{eq3}) the notation $dV = r_{32} r_{31} r_{21} dr_{32} dr_{31} dr_{21}$ is an elementary volume in the relative coordinates $r_{32}, r_{31}$ and $r_{21}$. Each of the relative coordinates 
is the difference between the two Cartesian coordinates of the corresponding particles, e.g., $r_{ij} = \mid {\bf r}_i - {\bf r}_j \mid$. It follows from such a definition that: (a) relative coordinates 
are symmetric, i.e. $r_{ij} = r_{ji}$, and (b) the following inequalities are always obeyed: $\mid r_{ik} - r_{jk} \mid \le r_{ij} \le r_{ik} + r_{jk}$ for $( i, j, k) = ( 1, 2, 3)$. Now, we can transform 
the expression in the right-hand side of Eq.(\ref{eq3}) to the following form
\begin{equation}
 M_{i \rightarrow f} = \int \phi_n(r_{31}) j_{0}(p r_{32}) \Bigl[\frac{\partial}{\partial r_{32}} \Phi_{{\rm H}^{-}}(r_{32}, r_{31}, r_{21}) \Bigr] dV 
 \; \; \; . \; \; \; \label{eq4}
\end{equation} 
which contains the partial derivatives of the incident wave function only, i.e. the bound state wave function of the H$^{-}$ ion. In the case of exponential variational, Eq.(\ref{exp}), expansion it is 
straightforward to derive the explicit analytical expression for the partial derivative in the last equation (all details can be found in \cite{Fro2015}). 

After a few additional steps of transformations one finds that calculations of the photodetachment amplitude $M_{i \rightarrow f}$ are reduced to analytical/numerical computations of the following 
integral which include one spherical Bessel function $j_{0}(p r_{32})$
\begin{eqnarray}
 I(\alpha, \beta, \gamma; p) &=& \int \int \int \exp(-\alpha r_{32} - \beta r_{31} - \gamma r_{21}) j_{0}(p r_{32}) r_{32} r_{31} r_{21} dr_{32} dr_{31} dr_{21} \nonumber \\
 &=& \frac{1}{p} \int \int \int \exp(-\alpha r_{32} - \beta r_{31} - \gamma r_{21}) \sin(p r_{32}) r_{31} r_{21} dr_{32} dr_{31} dr_{21} \label{j0}
\end{eqnarray} 
It is clear that analytical and/or numerical computations of three-particle integrals written in the relative coordinates is a difficult problem, since three relative coordinates $r_{32}, r_{31}$ and 
$r_{21}$ are not truly independent. Indeed, three inequalities $\mid r_{ik} - r_{jk} \mid \le r_{ij} \le r_{ik} + r_{jk}$ are always obeyed for three relative coordinates $r_{32}, r_{31}$ and $r_{21}$ 
(see above). In real three-body calculations it is better to use three perimetric coordinates $u_1, u_2, u_3$ which are simply related with the perimetric coordinates: $u_i = \frac12 (r_{ij} + r_{ik} 
- r_{jk})$, where $(i, j, k) = (1, 2, 3)$ and $r_{ij} = r_{ji}$. The inverse relations are: $r_{ij} = u_{i} + u_{j}$. Each of the three perimetric coordinates $u_i (i = 1, 2, 3)$ is: (1) non-negative, 
(2) truly independent from other perimetric coordinates, and (3) changes from zero to infinity. In perimetric coordinates the last integral from Eq.(\ref{j0}) is written in the form
\begin{eqnarray}
 I(\alpha, \beta, \gamma; p) &=& \frac{2}{p} \int^{+\infty}_{0} \int^{+\infty}_{0} \int^{+\infty}_{0} \exp[-(\alpha + \beta) u_3 - (\alpha + \gamma) u_{2} - (\beta + \gamma) u_{1}] \times \nonumber \\
 & & \sin(p u_{2} + p u_{3}) (u^{2}_{1} + u_{1} u_{2} +  u_{1} u_{3} +  u_{2} u_{3}) du_{1} du_{2} du_3 \label{j01}
\end{eqnarray} 
where the factor 2 is the Jacobian of the transition from the relative to perimetric coordinates, i.e. $(r_{32}, r_{31}, r_{21}) \rightarrow (u_1, u_2, u_3)$. To derive the final expression we apply the 
well known trigonometric formula $\sin(p u_{2} + p u_{3}) = \sin(p u_{2}) \cos(p u_{3}) + \cos(p u_{2}) \sin(p u_{3})$ and introduce the following notations: $Z = \alpha + \beta, Y = \alpha + 
\gamma, X = \beta + \gamma$. In these notations the integral Eq.(\ref{j01}) is written as the sum of eight simple integrals (all details are explained in \cite{Fro2015}) which are determined with 
the use of the formulas from \cite{GR}. The final expression for the integral $I(\alpha, \beta, \gamma; p)$ takes the form
\begin{eqnarray}
 I(\alpha, \beta, \gamma; p) &=& \frac{2}{X (Y^{2} + p^{2}) (Z^{2} + p^{2})} \Bigl[ \frac{2 (Z + Y)}{X^{2}} + \frac{2 Y Z}{X (Y^{2} + p^{2})} + \frac{2 Y Z}{X (Z^{2} + p^{2})} \label{j11} \\ 
 &+& \frac{Y^{2} - p^2}{X (Y^{2} + p^{2})} + \frac{Z^{2} - p^2}{X (Z^{2} + p^{2})} + \frac{2 Y (Z^{2} - p^2)}{(Y^{2} + p^{2}) (Z^{2} + p^{2})} + \frac{2 Z (Y^{2} - p^2)}{(Y^{2} + p^{2}) (Z^{2} + p^{2})}
 \Bigr] \nonumber 
\end{eqnarray} 
Note that the formula, Eq.(\ref{j11}), does not contain any singular and/or quasi-singular term represented by the Frullanian integrals \cite{Frul} which must be regularized before actual 
numerical computations. This is an obvious advantage of our approach which is based on the Rayleigh's formula, Eqs.(\ref{eq1}) - (\ref{eq2}) for spherical Bessel functions. 

The formula, Eq.(\ref{j11}), has been used in our numerical calculations performed in this study. The photodetachment cross-sections of the ${}^{\infty}$H$^{-}$ ion determined for different momenta of 
photo-electron $p_e$ can be found in Tables I - IV. Note that in this study we restrict ourselves to the consideration of the cases when the final hydrogen atom is formed in the ground $1s$-state. The 
energy conservation leads to the following formula $\hbar \omega = I_{1} + \frac{p^{2}_{e}}{2 m_e} = - E({\rm H}^{-}) + E(H, 1s) + \frac{p^{2}_{e}}{2 m_e}$. In atomic units this formula gives an important 
relation between frequency of light $\omega$ and momentum of the outgoing photo-electron $p_e$: $\omega = I_{1} + \frac{p^{2}_{e}}{2} = - E({\rm H}^{-}) + E(H, 1s) + \frac{p^{2}_{e}}{2}$ and $E(H, 1s) =
-0.5$ $a.u.$ (exactly). To the best of our knowledge this work is the first study where highly accurate wave functions are applied to determine the photodetachment cross-section(s) of the two-electron 
H$^{-}$ ion. The use of highly accurate variational wave functions for the negatively charged H$^{-}$ ion allows us to solve a number of long standing problems. First, we investigate the convergence of 
the photodetachment cross-section(s) upon the accuracy of the bound state wave functions used. In reality this means investigation of the $\sigma_{ph}(N)$ dependence, where $N$ is the total number of basis 
functions in the trial wave function of the H$^{-}$ ion used in calculations. This problem is analyzed in Table I where the photodetachment cross-sections of the H$^{-}$ ion are determined with $N$ = 2000, 
2500, 3000, 3500 and 4000 basis functions (exponents) in Eq.(\ref{exp}). The corresponding total energies $E$ obtained with the same wave functions of the H$^{-}$ ion are also presented in Table I. 

Another important question which was essentially ignored in previous studies is the location of the maximum of the photodetachment cross-section $\sigma_{ph}(p_e)$ of the H$^{-}$ ion. This question has 
been investigated by using our data from highly accurate computations (see Table II). Finally, we have found that the maximum of the photodetachment cross-section $[\sigma_{ph}(p_e)]_{\max} \approx$ 
3.862703449$\cdot 10^{-17}$ $cm^2$ is located at $p_e \approx$ 0.113206(1) $a.u.$ The location of this maximum does not depend upon the $p_e$ value, if the total number of basis functions (exponents) in 
Eq.(\ref{exp}) exceeds 1000. In other words, this maximum does not shift along the $p_e$-axis when the total number of basis functions $N$ increases (where $N \ge$ 1000). Another important result is the 
vanishing photoionization cross-section at $p_e = 0$, i.e. $\sigma_{ph}(0) = 0$. This directly follows from our calculations, since the code works for $p_e = 0$. Tables III and IV contain photodetachment 
cross-sections $\sigma_{ph}(p_e)$ of the H$^{-}$ ion determined for very small and very large values of the photo-electron momenta $p_e$. Photodetachment of the H$^{-}$ ion for these values of the 
photo-electron momenta $p_e$ has not been studied carefully in earlier studies. The results from Tables I - IV allow one to approximate them by using different representations for the $\sigma_{ph}(p_e)$ 
function. Then by using our computational data we can determine the coefficients $C_1, C_2, C_3, \ldots$ required in relatively simple interpolation formulas. In an ideal case the knowledge of a very few 
such coefficients allow one to approximate the exact computational results to high numerical accuracy. Some of such interpolation formulas for the photodetachment cross-sections of the H$^{-}$ ion will be 
discussed in our next study.  

It should be noted in conclusion that our results obtained in this study represent, in fact, the `theoretically final' solution of the photodetachment problem for the negatively charged hydrogen ion 
H$^{-}$ how it was formulated by Chandrasekhar in 1945 (see references in \cite{Chand2}). In \cite{Chand2}) the wave function of the outgoing (free) electron is represented as a Bessel function 
$j_{\ell=1}(p_e r)$. Also, it was assumed in \cite{Chand2}) that the final hydrogen atom H is formed in its ground $1^{2}s-$state. By evaluating the photodetachment cross-sections of the H$^{-}$ ion we
have noticed that the maximal deviation of our results presented in this work from the experimental results published in \cite{PRA2015} is less than 1\% - 2\% (exact deviation depends upon the momentum of 
photo-electron $p_e$). This is a good indication of the correctness of our method. In the next studies we are planning to generalize our method to the cases when the final hydrogen atom is formed in 
different excited states.

\newpage
\begin{table}[tbp]
   \caption{Photodetachment cross-section (in $cm^{2}$) of the negatively charged ${}^{\infty}$H$^{-}$ ion in the ground $1^1S-$state. 
            Convergence of the results upon the total number of basis functions $N$ in the trial wave function and $E$ is the total 
            energy (in $a.u.$) for this wave function. The final hydrogen atom is formed in the ground $1s$-state. 
            The notation $p_e$ stands for the absolute value of momentum of the emitted photo-electron (in $a.u.$).}
     \begin{center}
     \scalebox{0.55}{%
     \begin{tabular}{| c | c | c | c | c | c |}
      \hline\hline
 $E_{{\rm H}^{-}}$  & -0.52775101654437719641932 & -0.52775101654437719658149 & -0.52775101654437719658929 & -0.52775101654437719659013 & -0.52775101654437719659040 \\ 
           \hline
    $p_e$   & $\sigma(E_e)$ ($N$ = 2000)         & $\sigma(E_e)$ ($N$ = 2500)      & $\sigma(E_e)$ ($N$ = 3000)         & $\sigma(E_e)$ ($N$ = 3500)         & $\sigma(E_e)$ ($N$ = 4000) \\
          \hline    
  0.005 & 2.959858851387538$\cdot 10^{-18}$ & 2.959858874082872$\cdot 10^{-18}$ & 2.959858883655716$\cdot 10^{-18}$ & 2.959858883190412$\cdot 10^{-18}$ & 2.959858883004252$\cdot 10^{-18}$ \\
  0.015 & 8.795323872003963$\cdot 10^{-18}$ & 8.795323913839535$\cdot 10^{-18}$ & 8.795323937347148$\cdot 10^{-18}$ & 8.795323935993190$\cdot 10^{-18}$ & 8.795323935735443$\cdot 10^{-18}$ \\
  0.025 & 1.438345860752727$\cdot 10^{-17}$ & 1.438345861044433$\cdot 10^{-17}$ & 1.438345863600382$\cdot 10^{-17}$ & 1.438345863401517$\cdot 10^{-17}$ & 1.438345863421710$\cdot 10^{-17}$ \\
  0.035 & 1.957685162296752$\cdot 10^{-17}$ & 1.957685154067537$\cdot 10^{-17}$ & 1.957685155756557$\cdot 10^{-17}$ & 1.957685155553194$\cdot 10^{-17}$ & 1.957685155628780$\cdot 10^{-17}$ \\
  0.045 & 2.425225285598434$\cdot 10^{-17}$ & 2.425225268387473$\cdot 10^{-17}$ & 2.425225268797405$\cdot 10^{-17}$ & 2.425225268658815$\cdot 10^{-17}$ & 2.425225268736251$\cdot 10^{-17}$ \\
  0.055 & 2.831649718769707$\cdot 10^{-17}$ & 2.831649695942023$\cdot 10^{-17}$ & 2.831649695319822$\cdot 10^{-17}$ & 2.831649695282078$\cdot 10^{-17}$ & 2.831649695291207$\cdot 10^{-17}$ \\
  0.065 & 3.170944356423218$\cdot 10^{-17}$ & 3.170944332567042$\cdot 10^{-17}$ & 3.170944331479658$\cdot 10^{-17}$ & 3.170944331530498$\cdot 10^{-17}$ & 3.170944331444445$\cdot 10^{-17}$ \\
  0.075 & 3.440395691148654$\cdot 10^{-17}$ & 3.440395669680828$\cdot 10^{-17}$ & 3.440395668694068$\cdot 10^{-17}$ & 3.440395668801305$\cdot 10^{-17}$ & 3.440395668653597$\cdot 10^{-17}$ \\
  0.085 & 3.640333665730176$\cdot 10^{-17}$ & 3.640333648189376$\cdot 10^{-17}$ & 3.640333647715447$\cdot 10^{-17}$ & 3.640333647861971$\cdot 10^{-17}$ & 3.640333647714390$\cdot 10^{-17}$ \\
  0.095 & 3.773684860837691$\cdot 10^{-17}$ & 3.773684847532678$\cdot 10^{-17}$ & 3.773684847784903$\cdot 10^{-17}$ & 3.773684847971124$\cdot 10^{-17}$ & 3.773684847868560$\cdot 10^{-17}$ \\
             \hline 
  0.105 & 3.845412400264125$\cdot 10^{-17}$ & 3.845412391067079$\cdot 10^{-17}$ & 3.845412392043501$\cdot 10^{-17}$ & 3.845412392262530$\cdot 10^{-17}$ & 3.845412392211801$\cdot 10^{-17}$ \\
  0.115 & 3.861916372960420$\cdot 10^{-17}$ & 3.861916367485774$\cdot 10^{-17}$ & 3.861916368991185$\cdot 10^{-17}$ & 3.861916369213587$\cdot 10^{-17}$ & 3.861916369194255$\cdot 10^{-17}$ \\
  0.125 & 3.830456341416103$\cdot 10^{-17}$ & 3.830456338758827$\cdot 10^{-17}$ & 3.830456340500367$\cdot 10^{-17}$ & 3.830456340689317$\cdot 10^{-17}$ & 3.830456340678219$\cdot 10^{-17}$ \\
  0.135 & 3.758640152613815$\cdot 10^{-17}$ & 3.758640151270025$\cdot 10^{-17}$ & 3.758640152995945$\cdot 10^{-17}$ & 3.758640153135834$\cdot 10^{-17}$ & 3.758640153122914$\cdot 10^{-17}$ \\
  0.145 & 3.654004946057814$\cdot 10^{-17}$ & 3.654004944364808$\cdot 10^{-17}$ & 3.654004945957350$\cdot 10^{-17}$ & 3.654004946064539$\cdot 10^{-17}$ & 3.654004946051814$\cdot 10^{-17}$ \\
  0.155 & 3.523699997848411$\cdot 10^{-17}$ & 3.523699994708443$\cdot 10^{-17}$ & 3.523699996169406$\cdot 10^{-17}$ & 3.523699996274543$\cdot 10^{-17}$ & 3.523699996265742$\cdot 10^{-17}$ \\
  0.165 & 3.374268534173559$\cdot 10^{-17}$ & 3.374268529534476$\cdot 10^{-17}$ & 3.374268530890153$\cdot 10^{-17}$ & 3.374268531010272$\cdot 10^{-17}$ & 3.374268531003245$\cdot 10^{-17}$ \\
  0.175 & 3.211517380805054$\cdot 10^{-17}$ & 3.211517375556149$\cdot 10^{-17}$ & 3.211517376770000$\cdot 10^{-17}$ & 3.211517376895906$\cdot 10^{-17}$ & 3.211517376883291$\cdot 10^{-17}$ \\
  0.185 & 3.040458933894263$\cdot 10^{-17}$ & 3.040458929267448$\cdot 10^{-17}$ & 3.040458930231542$\cdot 10^{-17}$ & 3.040458930338886$\cdot 10^{-17}$ & 3.040458930313824$\cdot 10^{-17}$ \\ 
  0.195 & 2.865308631884608$\cdot 10^{-17}$ & 2.865308628763854$\cdot 10^{-17}$ & 2.865308629368181$\cdot 10^{-17}$ & 2.865308629439345$\cdot 10^{-17}$ & 2.865308629399930$\cdot 10^{-17}$ \\ 
             \hline 
  0.205 & 2.689521930504365$\cdot 10^{-17}$ & 2.689521929044216$\cdot 10^{-17}$ & 2.689521929262060$\cdot 10^{-17}$ & 2.689521929299735$\cdot 10^{-17}$ & 2.689521929249002$\cdot 10^{-17}$ \\ 
  0.215 & 2.515856856292309$\cdot 10^{-17}$ & 2.515856855995308$\cdot 10^{-17}$ & 2.515856855917814$\cdot 10^{-17}$ & 2.515856855941647$\cdot 10^{-17}$ & 2.515856855884585$\cdot 10^{-17}$ \\ 
  0.225 & 2.346450841574501$\cdot 10^{-17}$ & 2.346450841689923$\cdot 10^{-17}$ & 2.346450841487138$\cdot 10^{-17}$ & 2.346450841519598$\cdot 10^{-17}$ & 2.346450841460479$\cdot 10^{-17}$ \\ 
  0.235 & 2.182903235955107$\cdot 10^{-17}$ & 2.182903235917645$\cdot 10^{-17}$ & 2.182903235755902$\cdot 10^{-17}$ & 2.182903235809937$\cdot 10^{-17}$ & 2.182903235751758$\cdot 10^{-17}$ \\ 
  0.245 & 2.026357346601315$\cdot 10^{-17}$ & 2.026357346273983$\cdot 10^{-17}$ & 2.026357346243211$\cdot 10^{-17}$ & 2.026357346319736$\cdot 10^{-17}$ & 2.026357346265064$\cdot 10^{-17}$ \\ 
  0.255 & 1.877577931653412$\cdot 10^{-17}$ & 1.877577931298093$\cdot 10^{-17}$ & 1.877577931388224$\cdot 10^{-17}$ & 1.877577931482291$\cdot 10^{-17}$ & 1.877577931433757$\cdot 10^{-17}$ \\ 
  0.265 & 1.737021712279301$\cdot 10^{-17}$ & 1.737021712334526$\cdot 10^{-17}$ & 1.737021712465310$\cdot 10^{-17}$ & 1.737021712573886$\cdot 10^{-17}$ & 1.737021712533455$\cdot 10^{-17}$ \\ 
  0.275 & 1.604899698414829$\cdot 10^{-17}$ & 1.604899699233895$\cdot 10^{-17}$ & 1.604899699312594$\cdot 10^{-17}$ & 1.604899699437591$\cdot 10^{-17}$ & 1.604899699405215$\cdot 10^{-17}$ \\  
  0.285 & 1.481230994606315$\cdot 10^{-17}$ & 1.481230996287363$\cdot 10^{-17}$ & 1.481230996259806$\cdot 10^{-17}$ & 1.481230996405004$\cdot 10^{-17}$ & 1.481230996377949$\cdot 10^{-17}$ \\ 
  0.295 & 1.365888332293738$\cdot 10^{-17}$ & 1.365888334656114$\cdot 10^{-17}$ & 1.365888334527933$\cdot 10^{-17}$ & 1.365888334693180$\cdot 10^{-17}$ & 1.365888334666840$\cdot 10^{-17}$ \\ 
             \hline 
  0.305 & 1.258635930076193$\cdot 10^{-17}$ & 1.258635932757882$\cdot 10^{-17}$ & 1.258635932582879$\cdot 10^{-17}$ & 1.258635932760351$\cdot 10^{-17}$ & 1.258635932730187$\cdot 10^{-17}$ \\ 
  0.315 & 1.159160473639600$\cdot 10^{-17}$ & 1.159160476245647$\cdot 10^{-17}$ & 1.159160476094555$\cdot 10^{-17}$ & 1.159160476269728$\cdot 10^{-17}$ & 1.159160476233262$\cdot 10^{-17}$ \\ 
  0.325 & 1.067096081652166$\cdot 10^{-17}$ & 1.067096083882655$\cdot 10^{-17}$ & 1.067096083811019$\cdot 10^{-17}$ & 1.067096083967737$\cdot 10^{-17}$ & 1.067096083925470$\cdot 10^{-17}$ \\ 
  0.335 & 9.820441220030166$\cdot 10^{-18}$ & 9.820441237177885$\cdot 10^{-18}$ & 9.820441237470468$\cdot 10^{-18}$ & 9.820441238736451$\cdot 10^{-18}$ & 9.820441238285472$\cdot 10^{-18}$ \\ 
  0.345 & 9.035886935302027$\cdot 10^{-18}$ & 9.035886947443728$\cdot 10^{-18}$ & 9.035886948606493$\cdot 10^{-18}$ & 9.035886949539591$\cdot 10^{-18}$ & 9.035886949099369$\cdot 10^{-18}$ \\ 
  0.355 & 8.313085129244258$\cdot 10^{-18}$ & 8.313085137562631$\cdot 10^{-18}$ & 8.313085139227422$\cdot 10^{-18}$ & 8.313085139883229$\cdot 10^{-18}$ & 8.313085139485274$\cdot 10^{-18}$ \\ 
  0.365 & 7.647858593465322$\cdot 10^{-18}$ & 7.647858599497326$\cdot 10^{-18}$ & 7.647858601242023$\cdot 10^{-18}$ & 7.647858601732217$\cdot 10^{-18}$ & 7.647858601389784$\cdot 10^{-18}$ \\ 
  0.375 & 7.036131400292129$\cdot 10^{-18}$ & 7.036131405368821$\cdot 10^{-18}$ & 7.036131406865092$\cdot 10^{-18}$ & 7.036131407311102$\cdot 10^{-18}$ & 7.036131407018067$\cdot 10^{-18}$ \\ 
  0.385 & 6.473975546692632$\cdot 10^{-18}$ & 6.473975551639923$\cdot 10^{-18}$ & 6.473975552728388$\cdot 10^{-18}$ & 6.473975553220910$\cdot 10^{-18}$ & 6.473975552958628$\cdot 10^{-18}$ \\ 
  0.395 & 5.957642580998728$\cdot 10^{-18}$ & 5.957642586107051$\cdot 10^{-18}$ & 5.957642586795428$\cdot 10^{-18}$ & 5.957642587373716$\cdot 10^{-18}$ & 5.957642587121219$\cdot 10^{-18}$ \\
            \hline 
  0.405 & 5.483583511486363$\cdot 10^{-18}$ & 5.483583516665228$\cdot 10^{-18}$ & 5.483583517073762$\cdot 10^{-18}$ & 5.483583517727496$\cdot 10^{-18}$ & 5.483583517470176$\cdot 10^{-18}$ \\
  0.415 & 5.048459691737112$\cdot 10^{-18}$ & 5.048459696738597$\cdot 10^{-18}$ & 5.048459697027339$\cdot 10^{-18}$ & 5.048459697714612$\cdot 10^{-18}$ & 5.048459697448407$\cdot 10^{-18}$ \\
  0.425 & 4.649146862548893$\cdot 10^{-18}$ & 4.649146867161738$\cdot 10^{-18}$ & 4.649146867468807$\cdot 10^{-18}$ & 4.649146868139577$\cdot 10^{-18}$ & 4.649146867870248$\cdot 10^{-18}$ \\
  0.435 & 4.282734097205138$\cdot 10^{-18}$ & 4.282734101364571$\cdot 10^{-18}$ & 4.282734101771013$\cdot 10^{-18}$ & 4.282734102386391$\cdot 10^{-18}$ & 4.282734102125492$\cdot 10^{-18}$ \\
  0.445 & 3.946519037386446$\cdot 10^{-18}$ & 3.946519041190820$\cdot 10^{-18}$ & 3.946519041713753$\cdot 10^{-18}$ & 3.946519042255770$\cdot 10^{-18}$ & 3.946519042015653$\cdot 10^{-18}$ \\
  0.455 & 3.638000511502957$\cdot 10^{-18}$ & 3.638000515161112$\cdot 10^{-18}$ & 3.638000515767438$\cdot 10^{-18}$ & 3.638000516238821$\cdot 10^{-18}$ & 3.638000516028706$\cdot 10^{-18}$ \\
  0.465 & 3.354869386503532$\cdot 10^{-18}$ & 3.354869390250916$\cdot 10^{-18}$ & 3.354869390880426$\cdot 10^{-18}$ & 3.354869391297488$\cdot 10^{-18}$ & 3.354869391121609$\cdot 10^{-18}$ \\
  0.475 & 3.094998309693525$\cdot 10^{-18}$ & 3.094998313712837$\cdot 10^{-18}$ & 3.094998314300899$\cdot 10^{-18}$ & 3.094998314684001$\cdot 10^{-18}$ & 3.094998314541808$\cdot 10^{-18}$ \\
  0.485 & 2.856430841066595$\cdot 10^{-18}$ & 2.856430845436534$\cdot 10^{-18}$ & 2.856430845930395$\cdot 10^{-18}$ & 2.856430846295876$\cdot 10^{-18}$ & 2.856430846183557$\cdot 10^{-18}$ \\
  0.495 & 2.637370352396314$\cdot 10^{-18}$ & 2.637370357076961$\cdot 10^{-18}$ & 2.637370357443831$\cdot 10^{-18}$ & 2.637370357799609$\cdot 10^{-18}$ & 2.637370357712001$\cdot 10^{-18}$ \\
    \hline\hline
  \end{tabular}}
  \end{center}
  \end{table}
\newpage
\begin{table}[tbp]
   \caption{Photodetachment cross-section (in $cm^{2}$) of the negatively charged ${}^{\infty}$H$^{-}$ ion in the ground $1^1S-$state. 
            Area of $p_e$ values close to the maximum of the cross-sectionis shown, where $p_e$ is for the absolute value of momentum 
            of the emitted photo-electron (in $a.u.$). The final hydrogen atom is formed in the ground $1s$-state.}
     \begin{center}
     \scalebox{0.95}{%
     \begin{tabular}{| c | c | c | c | c | c |}
      \hline\hline
   $p_e$ & $\sigma_{ph}$ &  $p_e$ & $\sigma_{ph}$ &  $p_e$ & $\sigma_{ph}$ \\
          \hline    
  0.113180 & 3.862703286961577$\cdot 10^{-17}$ & 0.113200 & 3.862703441533454$\cdot 10^{-17}$ & 0.113220 & 3.862703398724123$\cdot 10^{-17}$ \\
  0.113181 & 3.862703299378898$\cdot 10^{-17}$ & 0.113201 & 3.862703444080294$\cdot 10^{-17}$ & 0.113221 & 3.862703391403474$\cdot 10^{-17}$ \\
  0.113182 & 3.862703311302625$\cdot 10^{-17}$ & 0.113202 & 3.862703446133688$\cdot 10^{-17}$ & 0.113222 & 3.862703383589529$\cdot 10^{-17}$ \\
  0.113183 & 3.862703322732763$\cdot 10^{-17}$ & 0.113203 & 3.862703447693643$\cdot 10^{-17}$ & 0.113223 & 3.862703375282295$\cdot 10^{-17}$ \\
  0.113184 & 3.862703333669322$\cdot 10^{-17}$ & 0.113204 & 3.862703448760169$\cdot 10^{-17}$ & 0.113224 & 3.862703366481781$\cdot 10^{-17}$ \\
  0.113185 & 3.862703344112308$\cdot 10^{-17}$ & 0.113205 & 3.862703449333271$\cdot 10^{-17}$ & 0.113225 & 3.862703357187993$\cdot 10^{-17}$ \\
  0.113186 & 3.862703354061729$\cdot 10^{-17}$ & 0.113206 & 3.862703449412958$\cdot 10^{-17}$ & 0.113226 & 3.862703347400939$\cdot 10^{-17}$ \\
  0.113187 & 3.862703363517592$\cdot 10^{-17}$ & 0.113207 & 3.862703448999236$\cdot 10^{-17}$ & 0.113227 & 3.862703337120627$\cdot 10^{-17}$ \\
  0.113188 & 3.862703372479904$\cdot 10^{-17}$ & 0.113208 & 3.862703448092114$\cdot 10^{-17}$ & 0.113228 & 3.862703326347064$\cdot 10^{-17}$ \\
  0.113189 & 3.862703380948674$\cdot 10^{-17}$ & 0.113209 & 3.862703446691599$\cdot 10^{-17}$ & 0.113229 & 3.862703315080257$\cdot 10^{-17}$ \\
               \hline
  0.113190 & 3.862703388923908$\cdot 10^{-17}$ & 0.113210 & 3.862703444797698$\cdot 10^{-17}$ & 0.113230 & 3.862703303320214$\cdot 10^{-17}$ \\
  0.113191 & 3.862703396405615$\cdot 10^{-17}$ & 0.113211 & 3.862703442410419$\cdot 10^{-17}$ & 0.113231 & 3.862703291066943$\cdot 10^{-17}$ \\            
  0.113192 & 3.862703403393801$\cdot 10^{-17}$ & 0.113212 & 3.862703439529769$\cdot 10^{-17}$ & 0.113232 & 3.862703278320451$\cdot 10^{-17}$ \\
  0.113193 & 3.862703409888474$\cdot 10^{-17}$ & 0.113213 & 3.862703436155756$\cdot 10^{-17}$ & 0.113233 & 3.862703265080745$\cdot 10^{-17}$ \\
  0.113194 & 3.862703415889642$\cdot 10^{-17}$ & 0.113214 & 3.862703432288387$\cdot 10^{-17}$ & 0.113234 & 3.862703251347834$\cdot 10^{-17}$ \\
  0.113195 & 3.862703421397312$\cdot 10^{-17}$ & 0.113215 & 3.862703427927670$\cdot 10^{-17}$ & 0.113235 & 3.862703237121723$\cdot 10^{-17}$ \\
  0.113196 & 3.862703426411491$\cdot 10^{-17}$ & 0.113216 & 3.862703423073613$\cdot 10^{-17}$ & 0.113236 & 3.862703222402422$\cdot 10^{-17}$ \\           
  0.113197 & 3.862703430932188$\cdot 10^{-17}$ & 0.113217 & 3.862703417726222$\cdot 10^{-17}$ & 0.113237 & 3.862703207189937$\cdot 10^{-17}$ \\
  0.113198 & 3.862703434959409$\cdot 10^{-17}$ & 0.113218 & 3.862703411885505$\cdot 10^{-17}$ & 0.113238 & 3.862703191484275$\cdot 10^{-17}$ \\
  0.113199 & 3.862703438493162$\cdot 10^{-17}$ & 0.113219 & 3.862703405551470$\cdot 10^{-17}$ & 0.113239 & 3.862703175285446$\cdot 10^{-17}$ \\
    \hline\hline
  \end{tabular}}
  \end{center}
  \end{table}
\newpage
\begin{table}[tbp]
   \caption{Photodetachment cross-section (in $cm^{2}$) of the negatively charged ${}^{\infty}$H$^{-}$ ion in the ground $1^1S-$state. 
            The area of small $p_e$ values is shown, where $p_e$ is for the absolute value of momentum of the emitted 
             photo-electron (in $a.u.$). The final hydrogen atom is formed in the ground $1s$-state.}
     \begin{center}
     \scalebox{0.95}{%
     \begin{tabular}{| c | c | c | c | c | c |}
      \hline\hline
   $p_e$ & $\sigma_{ph}$ &  $p_e$ & $\sigma_{ph}$ &  $p_e$ & $\sigma_{ph}$ \\
          \hline    
  0.0001 & 5.926787749053725$\cdot 10^{-20}$ & 0.0021 & 1.244363839315231$\cdot 10^{-18}$ & 0.0051 & 3.018910446904520$\cdot 10^{-18}$ \\
  0.0002 & 1.185355850944473$\cdot 10^{-19}$ & 0.0022 & 1.303592482841636$\cdot 10^{-18}$ & 0.0052 & 3.077953380070245$\cdot 10^{-18}$ \\
  0.0003 & 1.778029529262154$\cdot 10^{-19}$ & 0.0023 & 1.362817390831388$\cdot 10^{-18}$ & 0.0053 & 3.136987505144539$\cdot 10^{-18}$ \\
  0.0004 & 2.370698111025507$\cdot 10^{-19}$ & 0.0024 & 1.422038393679108$\cdot 10^{-18}$ & 0.0054 & 3.196012653769272$\cdot 10^{-18}$ \\
  0.0005 & 2.963359897434993$\cdot 10^{-19}$ & 0.0025 & 1.481255321804970$\cdot 10^{-18}$ & 0.0055 & 3.255028657644937$\cdot 10^{-18}$ \\
  0.0006 & 3.556013189735558$\cdot 10^{-19}$ & 0.0026 & 1.540468005655810$\cdot 10^{-18}$ & 0.0056 & 3.314035348531745$\cdot 10^{-18}$ \\
  0.0007 & 4.148656289227755$\cdot 10^{-19}$ & 0.0027 & 1.599676275706238$\cdot 10^{-18}$ & 0.0057 & 3.373032558250720$\cdot 10^{-18}$ \\
  0.0008 & 4.741287497278865$\cdot 10^{-19}$ & 0.0028 & 1.658879962459739$\cdot 10^{-18}$ & 0.0058 & 3.432020118684787$\cdot 10^{-18}$ \\
  0.0009 & 5.333905115334014$\cdot 10^{-19}$ & 0.0029 & 1.718078896449789$\cdot 10^{-18}$ & 0.0059 & 3.490997861779868$\cdot 10^{-18}$ \\
  0.0010 & 5.926507444927294$\cdot 10^{-19}$ & 0.0030 & 1.777272908240954$\cdot 10^{-18}$ & 0.0060 & 3.549965619545969$\cdot 10^{-18}$ \\
        \hline
  0.0011 & 6.519092787692878$\cdot 10^{-19}$ & 0.0031 & 1.836461828430003$\cdot 10^{-18}$ & 0.0081 & 4.785685150107651$\cdot 10^{-18}$ \\
  0.0012 & 7.111659445376139$\cdot 10^{-19}$ & 0.0032 & 1.895645487647011$\cdot 10^{-18}$ & 0.0082 & 4.844390872623464$\cdot 10^{-18}$ \\
  0.0013 & 7.704205719844760$\cdot 10^{-19}$ & 0.0033 & 1.954823716556470$\cdot 10^{-18}$ & 0.0083 & 4.903082766224887$\cdot 10^{-18}$ \\
  0.0014 & 8.296729913099852$\cdot 10^{-19}$ & 0.0034 & 2.013996345858387$\cdot 10^{-18}$ & 0.0084 & 4.961760664785085$\cdot 10^{-18}$ \\
  0.0015 & 8.889230327287063$\cdot 10^{-19}$ & 0.0035 & 2.073163206289395$\cdot 10^{-18}$ & 0.0085 & 5.020424402268308$\cdot 10^{-18}$ \\
  0.0016 & 9.481705264707691$\cdot 10^{-19}$ & 0.0036 & 2.132324128623859$\cdot 10^{-18}$ & 0.0086 & 5.079073812730950$\cdot 10^{-18}$ \\
  0.0017 & 1.007415302782979$\cdot 10^{-18}$ & 0.0037 & 2.191478943674975$\cdot 10^{-18}$ & 0.0087 & 5.137708730322619$\cdot 10^{-18}$ \\
  0.0018 & 1.066657191929927$\cdot 10^{-18}$ & 0.0038 & 2.250627482295878$\cdot 10^{-18}$ & 0.0088 & 5.196328989287202$\cdot 10^{-18}$ \\ 
  0.0019 & 1.125896024195103$\cdot 10^{-18}$ & 0.0039 & 2.309769575380745$\cdot 10^{-18}$ & 0.0089 & 5.254934423963931$\cdot 10^{-18}$ \\
  0.0020 & 1.185131629882000$\cdot 10^{-18}$ & 0.0040 & 2.368905053865898$\cdot 10^{-18}$ & 0.0090 & 5.313524868788439$\cdot 10^{-18}$ \\
    \hline\hline
  \end{tabular}}
  \end{center}
  \end{table}
\newpage
\begin{table}[tbp]
   \caption{Photodetachment cross-section (in $cm^{2}$) of the negatively charged ${}^{\infty}$H$^{-}$ ion in the ground $1^1S-$state. 
            `Asymptotic' area of large $p_e$ values is shown, where $p_e$ is for the absolute value of momentum of the emitted 
             photo-electron (in $a.u.$). The final hydrogen atom is formed in the ground $1s$-state.}
     \begin{center}
     \scalebox{0.95}{%
     \begin{tabular}{| c | c | c | c | c | c |}
      \hline\hline
   $p_e$ & $\sigma_{ph}$ &  $p_e$ & $\sigma_{ph}$ &  $p_e$ & $\sigma_{ph}$ \\
          \hline    
  0.505 & 2.43616897649114$\cdot 10^{-18}$ & 0.705 & 5.494982965523706$\cdot 10^{-19}$ & 0.905 & 1.485335455039921$\cdot 10^{-19}$ \\
  0.515 & 2.25131677471321$\cdot 10^{-18}$ & 0.715 & 5.125867702767823$\cdot 10^{-19}$ & 0.915 & 1.397292838562950$\cdot 10^{-19}$ \\
  0.525 & 2.08143134554154$\cdot 10^{-18}$ & 0.725 & 4.783722635361872$\cdot 10^{-19}$ & 0.925 & 1.314971999351696$\cdot 10^{-19}$ \\
  0.535 & 1.92524785900944$\cdot 10^{-18}$ & 0.735 & 4.466432624341960$\cdot 10^{-19}$ & 0.935 & 1.237970047225239$\cdot 10^{-19}$ \\
  0.545 & 1.78160965759688$\cdot 10^{-18}$ & 0.745 & 4.172058397433862$\cdot 10^{-19}$ & 0.945 & 1.165914510785246$\cdot 10^{-19}$ \\
  0.555 & 1.64945942280800$\cdot 10^{-18}$ & 0.755 & 3.898821238838574$\cdot 10^{-19}$ & 0.955 & 1.098460897237527$\cdot 10^{-19}$ \\
  0.565 & 1.52783092457509$\cdot 10^{-18}$ & 0.765 & 3.645089053901800$\cdot 10^{-19}$ & 0.965 & 1.035290458477276$\cdot 10^{-19}$ \\
  0.575 & 1.41584135048405$\cdot 10^{-18}$ & 0.775 & 3.409363684021422$\cdot 10^{-19}$ & 0.975 & 9.761081452134533$\cdot 10^{-20}$ \\
  0.585 & 1.31268420123730$\cdot 10^{-18}$ & 0.785 & 3.190269358117466$\cdot 10^{-19}$ & 0.985 & 9.206407325764344$\cdot 10^{-20}$ \\
  0.595 & 1.21762273141292$\cdot 10^{-18}$ & 0.795 & 2.986542177108017$\cdot 10^{-19}$ & 0.995 & 8.686351021650597$\cdot 10^{-20}$ \\
        \hline
  0.605 & 1.12998390969667$\cdot 10^{-18}$ & 0.805 & 2.797020537139343$\cdot 10^{-19}$ & 1.05  & 6.348142813406855$\cdot 10^{-20}$ \\
  0.615 & 1.04915286976346$\cdot 10^{-18}$ & 0.815 & 2.620636405852337$\cdot 10^{-19}$ & 1.07  & 5.678360058813787$\cdot 10^{-20}$ \\
  0.625 & 9.74567821402757$\cdot 10^{-19}$ & 0.825 & 2.456407373777134$\cdot 10^{-19}$ & 1.09  & 5.085888396564905$\cdot 10^{-20}$ \\
  0.635 & 9.05715390952375$\cdot 10^{-19}$ & 0.835 & 2.303429410081573$\cdot 10^{-19}$ & 1.11  & 4.815572922397388$\cdot 10^{-20}$ \\
  0.645 & 8.42126360342966$\cdot 10^{-19}$ & 0.845 & 2.160870258405509$\cdot 10^{-19}$ & 1.13  & 4.095536099646251$\cdot 10^{-20}$ \\
  0.655 & 7.83371774840613$\cdot 10^{-19}$ & 0.855 & 2.027963414439222$\cdot 10^{-19}$ & 1.15  & 3.682034831932177$\cdot 10^{-20}$ \\
  0.665 & 7.29059390737712$\cdot 10^{-19}$ & 0.865 & 1.904002632296050$\cdot 10^{-19}$ & 1.18  & 3.145782445819049$\cdot 10^{-20}$ \\
  0.675 & 6.78830435652465$\cdot 10^{-19}$ & 0.875 & 1.788336911630659$\cdot 10^{-19}$ & 1.20  & 2.836547855893701$\cdot 10^{-20}$ \\
  0.685 & 6.32356655656815$\cdot 10^{-19}$ & 0.885 & 1.680365921906172$\cdot 10^{-19}$ & 1.25  & 2.200831704061166$\cdot 10^{-20}$ \\
  0.695 & 5.89337625087681$\cdot 10^{-19}$ & 0.895 & 1.579535824254425$\cdot 10^{-19}$ & 1.29  & 1.805282055766307$\cdot 10^{-20}$ \\
    \hline\hline
  \end{tabular}}
  \end{center}
  \end{table}
\end{document}